\documentclass{aa}  
\usepackage{graphicx}\usepackage[authoryear]{natbib}
\bibpunct{(}{)}{;}{a}{}{,}
\usepackage{txfonts}
\begin{document}
   \title{Near-infrared survey of High Mass X-ray Binary candidates}


   \author{J.M. Torrej\'on
          \inst{1,2}
          \and
          I. Negueruela
	  \inst{1}
	  \and
	  D. M. Smith
	  \inst{3}
          \and 
          T. E. Harrison
          \inst{4}
          }

   \offprints{J.M. Torrej\'on}

   \institute{Departamento de F\'{\i}sica, Ingenier\'{\i}a de Sistemas y Teor\'{\i}a de la Se\~nal, Universidad de Alicante, E-03080 Alicante, Spain\\
              \email{jmt@ua.es}
         \and
            Massachusetts Institute of Technology, Kavli Institute for Astrophysics and Space Research, Cambridge MA 02139\\
	 \and 
	 Physics Department and Santa Cruz Institute for Particle
  Physics, University of California, Santa Cruz, 1156 High St., Santa
  Cruz, CA 95064
  \and
Astronomy Department, New Mexico State University, Box
30001/Department 4500, Las Cruces, NM 88003
             }

   \date{Received ; accepted }

 
  \abstract
  { The \emph{INTEGRAL} satellite is discovering a large population of
    new X-ray sources which were missed by previous missions due to
    high obscuration and, in some cases, very short duty cycles. The
    nature of these sources must be addressed via the characterization
    of their optical and/or infrared counterparts. }
   {We investigate the nature of the optical counterparts to five of these newly discovered X-ray sources.}
   {We combine infrared spectra in the $I$, $J,H$ and $K$ bands
     together with $JHK$ photometry to characterize the spectral
     type, luminosity class and distance to the infrared counterparts
     to these systems. For IGR~J19140$+$0951, we present spectroscopy
     from the red to the $K$ band and new red and infrared
     photometry. For SAX~J18186$-$1703 and IGR~J18483$-$0311, we
     present the first intermediate-resolution spectroscopy
     reported. Finally, for IGR~J18027$-$2016, we present new $I$ and
     $K$ band spectra.}
   {We find that four systems harbour early-type B supergiants. All of
     them are heavily obscured, with $E(B-V)$ ranging between 3 and 5,
     implying visual extinctions of $\sim$ 9 to 15 magnitudes. We
     refine the published classifications of IGR~J18027$-$2016 and
     IGR~J19140$+$0951 by constraining their luminosity class. In the
     first case, we confirm the supergiant nature and rule out class
     III. In the second case, we propose a slightly higher luminosity
     class (Ia instead of Iab) and give an improved value of the
     distance based on new optical photometry. Two other systems,
     SAX~J18186$-$1703 and IGR~J18483$-$0311 are classified as 
     Supergiant Fast X-ray Transients (SFXTs). XTE~J1901$+$014, on the
   other hand, contains no bright infrared source in its error circle.}
   {Owing to their infrared and X-ray characteristics,
     IGR~J18027$-$2016 and IGR~J19140$+$0951, emerge as 
     Supergiant X-ray binaries with X-ray luminosities of the order of
     $L_{X}\sim [1-2]\times 10^{36}$ erg s$^{-1}$, while
     SAX~J1818.6$-$1703 and IGR~J18483$-$0311, turn out to be SFXTs at
     2 and 3 kpc, respectively. Finally, XTE~J1901$+$014 emerges as a
     puzzling source: its X-ray behaviour is strongly reminiscent of
     the SFXTs but a supergiant nature is firmly ruled out for the
     counterpart. We discuss several alternative scenarios to explain
     its behaviour. } 

   \keywords{Stars: binaries - supergiants - infrared: stars - X-rays: binaries
               }

   \maketitle
%

\section{Introduction}

The {\it INTEGRAL} satellite has broadened our knowledge of X-ray
binaries, by discovering a significant population of new X-ray
sources.  The Third {\it IBIS/ISGRI} cataloge \citep{bird07} contains more than 420 sources of which 167 are new.  The on-line cataloge of {\it INTEGRAL} sources\footnote{\tiny{\texttt{http://isdcul3.unige.ch/$\sim$rodrigue/html/igrsources.html}}}
 lists 245 sources discovered or re-discovered by {\it INTEGRAL}. Of these, 61 ($\sim 25$ \%) remain unidentified and 81 ($\sim 33$ \%) are extragalactic objects (AGNs, Seyferts, QSOs, etc).  The rest are mainly X-ray binaries. 51 of them ($\sim 21$ \%) turn out to be High Mass X-ray Binary candidates (HMXB),
which were missed by previous missions because of their high
obscuration, because of their very short X-ray cycles or a combination
of both. Approximately half of them have been confirmed so far. Amongst these new 51 HMXB systems, there are a few Be/X-ray binaries (BeXB),
Supergiant X-ray binaries (SGXB) and the newly established
class of Supergiant Fast X-ray Transients \citep[SFXT;
e.g.,][]{smith06,neg06a,sguera06}. This last class comprises 16 candidates of which nearly half have been confirmed.  Before we can conclusively classify every new source and understand its X-ray behavior, we must first establish the nature of the counterpart.

Since the majority of these sources are
located in the galactic plane, and concentrated towards the galactic center
region, they suffer from large extinction ($A_{V}\lesssim 15$ mag). Thus, their
detection in the visual is often very difficult, and other strategies must
be adopted. Detection and study of the counterparts in the infrared,
however, is possible with 4~m class telescopes. An important caveat,
though, is that the spectral classification of hot stars
using infrared spectra is much more uncertain than using the well
established MK standard 3950--4750 \AA\ region. For example, the
spectral classification based on the $K$ band spectrum cannot be done
without fundamental ambiguities because of the lack of adequate spectral
features in that range \citep{hanson96}. This problem can be, at least
partially, circumvented by using the combined information of several
spectral bands. However, this is often impossible to achieve within a
single observing campaign.   

Following the discovery by \emph{INTEGRAL}, intense observing
campaigns have been undertaken \citep[amongst
others]{chaty08,masetti08,nespoli08,neg09}. As these discoveries proceed,
it is fundamental to refine and establish the nature of these new
sources. In particular, a very important parameter is the distance to
the system which, in turn, gives the X-ray luminosity, the main
observable used to constrain the
theoretical models.  Most of the work done, however, has
been based on $H$ and/or $K$ band spectra to overcome the high
extinction. Classification based on a single infrared spectrum (in many
cases of intermediate or low resolution only) allows only a crude
approximation. For example, the difference in $M_{V}$ between a B1\,Ib
and a B1\,Ia supergiant (which are difficult or even impossible to
tell apart based on an intermediate resolution $H$ or $K$ spectrum)
results in an uncertainty of an order of magnitude error in $L_{X}$
\citep{neg09}. In this paper, we combine new spectroscopy in several bands
with new $JHK$ photometry (also $RI$ for
one source) to further refine the spectral classification of a sample
of these sources.  

\section{The observations}

\subsection{Photometry}

    \begin{table*}
      \caption[]{$JHK^{\prime}$ infrared photometry for the sources included in our survey. The corresponding {\it 2MASS} counterparts and the reference of their discovery are also included. }
 \label{tab:photometry}        
 	\begin{center}   
         \begin{tabular}{lccclccc}
            \hline
	    \hline
            \noalign{\smallskip}
            Source      &  2MASS    &  R.A. &  Dec. & Ref.  & $J$ ($1.27\mu$m)  &  $H$ ($1.63\mu$m) &  $K^{\prime}$ ($2.12 \mu$m) \\
            \noalign{\smallskip}
            \hline
            \noalign{\smallskip}
        IGR~J18027$-$2016 & 2MASS~J18024194$-$2017172 & 18 02 41.94  & $-$20 17 17.3 & C &12.79$\pm 0.06$ & 11.96$\pm 0.08$ & 11.48$\pm 0.05$\\
        SAX~J18186$-$1703 & 2MASS~J18183790$-$1702479 & 18 18 37.90  & $-$17 02 47.9 & Z1 & 10.08$\pm 0.06$ & 8.94$\pm 0.08$  & 7.79$\pm 0.05$   \\
        IGR~J18483$-$0311$^{a}$ & 2MASS~J18481720$-$0310168 & 18 48 17.20  & $-$03 10 16.8 & C & 10.84$\pm 0.03$ & 9.38$\pm 0.02$  & 8.47$\pm 0.02$\\
        IGR~J19140$+$0951 & 2MASS~J19140422$+$0952577 & 19 14 04.23  & +09 52 57.7 & Z2 & 11.33$\pm 0.06$ & 9.71$\pm 0.08$  & 8.83$\pm 0.05$ \\
            \noalign{\smallskip}
            \hline
         \end{tabular}
\begin{list}{}{}
\item C: \citet{chaty08}
\item Z1: \citet{intzand06a}
\item Z2: \citet{intzand06b}
\item [$^{\mathrm{a}}$]Photometry from \citet{rahoui08}. For this source, the filters are centered at
  1.25, 1.65 and $2.20\mu$m respectively. 
\end{list}
\end{center}
    \end{table*} 

$JHK^{\prime}$ band photometry was obtained with the 3.5~m Telescopio
Nazionale Galileo (TNG) at the Island of La Palma (Canary Islands,
Spain) during the nights of 2006 July 7 and 2007 May 1. In both runs
the configuration was the same. The instrument used was the Near
Infrared Camera and Spectrograph (NICS) in the Large Field
configuration, equipped with a Rockwell 1024$\times$1024 HgCdTe Hawaii
array detector, which gives a spatial scale of $0\farcs25$/pixel and
covers a field of view of 
$4\farcm2\times4\farcm2$. During the night of 2006 July 7, the seeing
was $0\farcs7$ in the $J$ band, very stable during the whole night,
while on 2007 May 1, the seeing was $0\farcs8$. 

Each observation consists of a mosaic of five images constructed from
a dithering pattern which kept the frame always on-source, using an
automatic script available at the telescope. From this mosaic,
flat-field correction, bad pixel map, sky
subtraction and image co-addition where performed with the {\sc SNAP}
software\fnmsep\footnote{\tt{http://www.tng.iac.es/news/2002/09/10/snap/index.html}}.  
The photometry was subsequently performed using Starlink {\sc gaia} on the 
final co-added frame.   

The Arnica standard star fields AS33 and AS27 \citep{hunt98} were observed
throughout the night at a range of airmasses. Transformation equations
were applied to the program star giving the magnitudes listed in
Table~\ref{tab:photometry}. 


Comparison with previously published values show some discrepancies
which can be readily explained. In particular, for IGR~J19140$+$0951,
comparison with 2MASS magnitudes shows a discrepancy. This is
due to the close proximity of the bright star 2MASS~J19149417+0952538,
located very close to the south-east, which contaminates the
  2MASS photometry \citep[cf.][]{intzand06b}. Any photometry made with a
3.5~m telescope under 
good seeing conditions must certainly be contaminated as well. However,
careful attention has been paid here to the extraction to minimize
this effect, making the magnitudes listed above rather more reliable than
those of the 2MASS catalogue. 


For IGR~J19140$+$0951, $RI$ band photometry was obtained
with the Andalucia Faint Object Spectrograph and Camera (ALFOSC) on
the 2.6~m Nordic Optical Telescope (NOT) in La Palma, on the night of
2006 June 18. The instrument was equipped with the thinned
2048$\times$2048 pixel E2V CCD, covering a field of view of
$6\farcm4\times6\farcm4$ with a spatial scale 0.19 arcsec/pixel. Standard
Bessel $R$ an interference $i$ filters were used to observe both the
target and the standard star fields. We obtain  $R=20.77\pm0.05$ and
$I=17.30\pm0.03$.

\subsection{Spectroscopy}

Infrared spectroscopy of several sources was performed with
NICS. Medium resolution infrared spectroscopy in the $K$ band was also 
obtained on 2006 July 7. The instrument 
was equipped with the $K_{\rm b}$ grism, which covers the
1.95--2.34~$\mu$m spectral region ,
and gives a resolution of 4.3\AA/pixel.  A slit of $1\arcsec$ was
chosen, providing a resolving power $R=1250$. 

To estimate the sky, we observed the target stars integrating on four
positions along the slit. We used the {\it ABBA} script which automatically
moves the telescope, nodding the source along the slit, thereby producing a set of four images with typical exposure times ranging from 100 to 300 s for the programme stars thereby minimising the sky variability throughout each observation. The four images
were median averaged and this median, subtracted from each individual
image of the set. This method, together with the use of the $1\arcsec$ slit, also minimizes any possible nebular contamination.  These sky subtracted images were flat fielded and
the resulting spectra averaged. The wavelength calibration was done
with the Ar lamps available at the telescope. Analysis was subsequently performed with Starlink \textsc{figaro} and \textsc{dipso}.

A number of A0\,V and G2\,V stars were selected from
the list available at the telescope and observed throughout the night
at airmasses and positions as close as possible to the target fields.  

For the telluric subtraction two methods were used. The first one,
described in \citet{hanson96}, makes use of the fact that spectra of
A0\,V stars are featureless in the $K$ band except for the Br$\gamma$
line. A spectrum from the Sun was retrieved from the ESO
archives. This spectrum was rebinned to match the sampling of our
target spectra, wavelength shifted for the different radial velocities
and, finally, the lines rotationally broadened to match the profiles
of our G2\,V standard stars. The spectrum of our G2\,V stars was then
divided by this corrected spectrum of the Sun. The resulting spectrum
was subsequently used to correct the Br$\gamma$ zone in the spectrum
of the A0\,V standard star, isolating the true Br$\gamma$ profile of the
early type star. Then, we fitted this line only and removed it from
the A0\,V spectrum, leaving a pure telluric spectrum which was
subsequently applied to the programme stars.  

The second method is described in
\citet{maiolino}. Essentially, it uses directly the G2\,V star spectrum
divided by the modified solar spectrum to correct the whole spectrum of the
program star. At the resolution of our spectra, these two methods give
almost indistinguishable results. Where appropriate, we averaged the
final spectra, of each source, obtained by both methods. The resulting
spectra have been normalised to unity to allow for comparison with
standard atlases. 
   
In addition, infrared spectroscopy of the counterpart to IGR~J19140$+$0951 was obtained using SPEX on the Infrared Telescope
Facility (IRTF) on Mauna Kea on 2005 September 3. The $J$, $H$ and $K$ band
spectra presented in Fig. \ref{fig:19140_spec} were constructed from medians of six
10~s exposures. The observing and data-reduction procedure are fully 
described in \citet{harrison04}. SPEX was used in single-order
mode with a $0\farcs3$ slit, giving a dispersion of 5.51
\AA/pixel. Unfortunately, the conditions at the IRTF were not 
photometric, but the seeing was excellent.      

     We employed a script with which data at six separate positions
     along the slit were obtained. To remove the sky background and
     dark current from each SPEX exposure, we subtracted the median of
     the other five exposures obtained in an observing sequence. This
     process resulted in six background-subtracted exposures from
     which the spectra were extracted using the normal IRAF
     methods. The spectra were wavelength calibrated by the extraction
     of an Ar arc spectrum at the position (aperture) of each
     spectrum. We observed the G2\,V star HD~150698 (in an identical
     fashion) to correct for telluric features as described above.

Finally, $I$-band spectra of  IGR~J18027$-$2016 and  IGR~J18483$-$0311
were obtained on 2008 May 16 using the 4.2-m William Herschel
Telescope (WHT), in La Palma (Spain) equipped with the ISIS double-beam
spectrograph, during a service run. We used the red arm fitted with
the R316R grating and the RED$+$ CCD, a configuration that results
in a nominal dispersion of 0.85\AA/pixel (the resolution element is
approximately 3 pixels). We also observed the counterpart
to IGR~J19140$+$0951 on 2008 May 18 using the same instrument, but this
time fitted with the R158R grating, which results in a nominal
dispersion of 1.8~\AA/pixel.

\section{Results}
  
   For the line identification and spectral classification we have
   used the following atlases: \citet{andrillat95} for the $I$ band, \citet{wallace00} for the $J$ band;
   \citet{blum97}, \citet{meyer98} and \citet{hanson98} for the $H$ band;
   \citet{hanson96} and \citet{hanson05} for the $K$ band. 

 As has been pointed out, the classification of massive stars using the 2 micron spectral range is difficult due to the lack of spectral features in this domain. In the case of HMXB the situation is still worse because the spectra of the infrared counterparts might be altered by the presence of the compact object in the system. Little work has been done in the near-infrared observations of HMXBs (see, for example, \cite{clark99}; Fig. 7, for some BeXs). In the pioneering work by \cite{hanson96} a number of HMXB are included (cf. Fig. 29 in that reference). In almost all of them, the relevant diagnostic lines appear in emission, both in BeXs and in SGXBs. However, isolated supergiants tend to show these lines in absorption \citep{hanson96}. No HMXBs are included in the higher S/N survey of \cite{hanson05}. Notwithstanding, this last work has been used already to classify the infrared counterparts of some HMXBs \citep[vg.]{mason09}. Lacking a comprehensive scheme to classify HMXB counterparts from near-infrared spectra, we will use the previously cited available atlases. We estimate that the uncertainty due to the above mentioned effects alone could amount to half a spectral subtype. 

In Table \ref{tab:ew} we give the equivalent widths of the main diagnostic lines found in our spectra. Due caution must be exercised, however, in using them for spectral classification purposes, as stated by \citet[][;see their Section 5.2]{hanson05} as only the comparison between different classes appears to be meaningful.

    \begin{table*}
      \caption[]{Equivalent widths (\AA) of the main diagnostic lines in the $JHK$ bands observed in our sample. Typical errors are always less 1 \AA. Numbers beside chemical species are the laboratory wavelengths in $\mu$m. $I$ band equivalent widths have not been measured due to the line blending. }
 \label{tab:ew}        
 	\begin{center}   
         \begin{tabular}{ccccccccc}
            \hline
	    \hline
            \noalign{\smallskip}
            Source      &  \ion{He}{i}~2.058$\mu$m   & \ion{He}{i}~2.1126  &  \ion{N}{iii}~2.1155 & \ion{Br}{$\gamma$}~2.1661 & \ion{He}{i}~1.083 &  \ion{Pa}{$\beta$}~1.2818 &  \ion{Br}{11}~1.6814 & \ion{He}{i}~1.7004 \\
            \noalign{\smallskip}
            \hline
            \noalign{\smallskip}
        IGR~J18027$-$2016 & 4.1(a+e) & 2.6 & -0.2 &  1.1 (a+e)& ... & ... & ... & ...\\
                                         & -4.5(e)   &          &           & -4.1 (e)  &      &     &     &    \\
                                         &                    &            &     &                  &                   &               &              &     \\
        SAX~J18186$-$1703 & 8.6(a+e)  & 2.0 & ... & 5.1 & ... & ...  & ...   & ... \\
                                          & -6.5(e)   &          &     &        &      &      &       &     \\
                                         &                    &            &     &                  &                   &               &              &     \\
        IGR~J19140$+$0951 $^{\dag}$ & -1.9(a+e) & 2.2 & ... & 2.0(a+e) & ... & ... & ... & ...\\
                                          & -5.7(e)      &            &     & -1.0(e)    &          &     &         &      \\
                                          &                    &            &     &                  &                   &               &              &     \\
        IGR~J19140$+$0951 $^{\ddag}$  & -1.1(a+e)  &  2.2   & -0.4 & 1.2(a+e) & -7.1   & 1.8(a+e)   & 2.3 &  2.1\\
                                          & -1.3(e)      &            &           &  -2.6(e)  &    & -5.3(e)    &   &  \\
            \noalign{\smallskip}
            \hline
         \end{tabular}
\begin{list}{}{}
\item (a+e): Absorption + Emission;  (e): Emission only
\item [$^{\dag}$]TNG data;  $^{\ddag}$ SPEX-IRTF data
\end{list}
\end{center}
    \end{table*}

 In order to compute their distance and $L_{X}$
we will use the infrared distance modulus computed from the following
equation.

\begin{equation}
K_{0}-M_{K}=5\log d-5 \ ,
\end{equation}

where $K_{0}=K-A_{K}$ are the infrared magnitudes corrected for
extinction and $K$ magnitudes are taken from Table
\ref{tab:photometry}. We compute the $M_{K}$ band absolute
magnitudes from the $(V-K)_{0}$ colors given by \citet{ducati01}
taking into account the $M_{V}$ values from \citet{sk82}. To compute
the total to selective 
absorption $A$ we will first compute the corresponding IR color excess
$E(J-K)=(J-K)-(J-K)_{0}$, where the intrinsic IR colors are taken from
\citet{ducati01} for the deduced spectral type. These excesses are converted to $E(B-V)$
by means of the following relation $E(J-K)=0.50E(B-V)$. Finally
$A_{K}=0.36E(B-V)$ \citep{fitzpatrick}, assuming the standard extinction law ($R_{\rm V}=3.1$). We compute the
distance for the two extremes of the range in spectral type compatible
with our observations, namely,
the cold-faint and hot-luminous ends. The range of possible distances
so deduced and the corresponding X-ray luminosities are given in
Table~\ref{tab:data}. The quoted errors refer only to the dispersion of these estimated values. The final uncertainties in the distance will be dominated by the intrinsic dispersion of the calibration of the order of 0.5 mag.

\subsection{IGR~J18027$-$2016}

\begin{figure*}
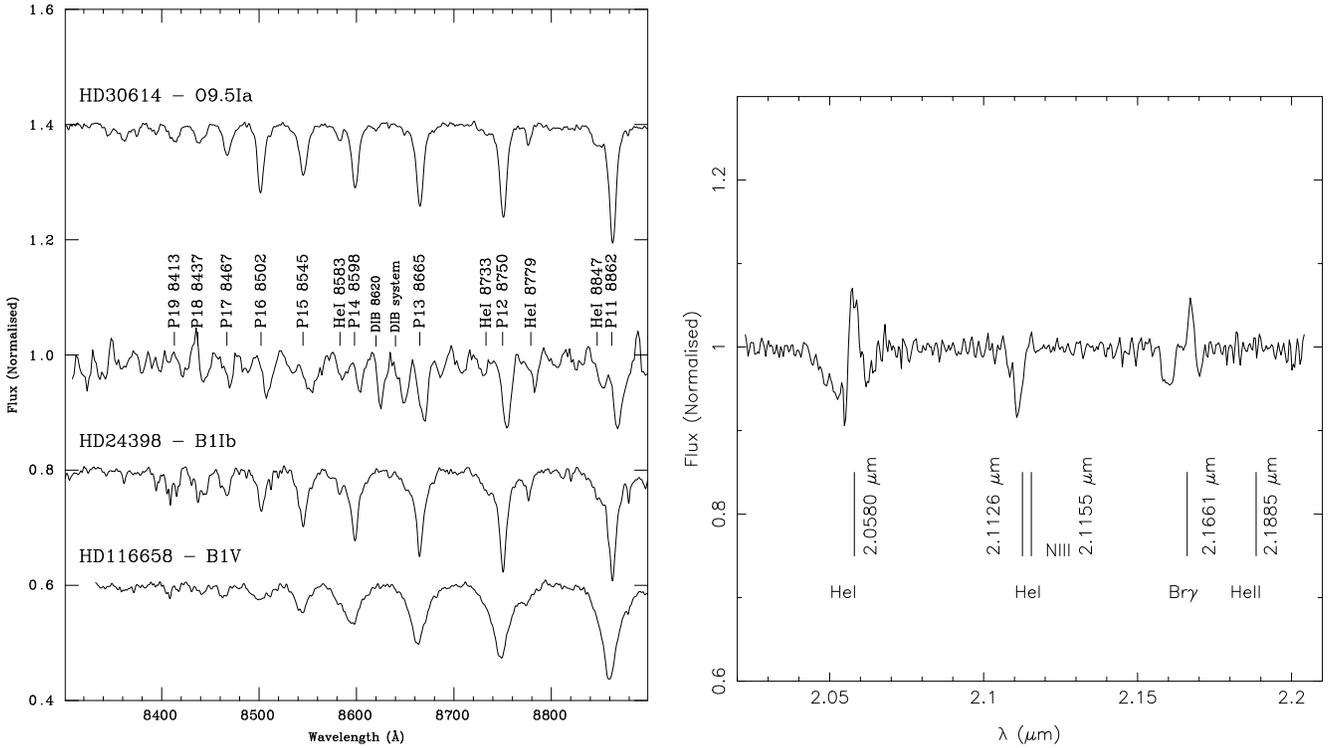

   \centering
   \includegraphics[width=\columnwidth]{igrj1802.eps}
   \includegraphics[width=\columnwidth,clip]{18027_K.ps}
   \caption{  {\bf Left:} {\it WHT} $I$-band spectrum of IGR~J18027$-$2016 normalised to flux unity bracketted between comparison standard stars.  {\bf Right:} {\it TNG} $K$-band spectrum normalised to flux unity. Emission is clearly seen in
     Br$\gamma$ and \ion{He}{i} lines, and also likely in
     \ion{N}{iii}~2.1155$\mu$m, suggesting an early B supergiant. }
    \label{fig:18027_both_spec}
\end{figure*}  

This source was discovered by \citet{revniv04} using
{\it INTEGRAL}. The source has a very large hydrogen column ($N_{\rm H}=(6.8\pm 1)\times
10^{22}\:{\rm cm}^{-2}$). It has been identified with SAX~J1802.7$-$2017, 
which was discovered serendipitously by \citet{augello03} with {\it
  BeppoSAX}. Using {\it XMM-Newton}, \citet{walter06} localised
the source position at $\alpha=18\:{\rm h}$~$02\:{\rm m}$~$42.0\:{\rm
  s}$, $\delta=-20\degr\:17\arcmin\: 18\arcsec$ with a $4\arcsec$
positional accuracy. Its 
2--10~keV unabsorbed flux during the primary pulse is  $8.9\times
10^{-11}\:{\rm erg}\,{\rm s}^{-1}\,{\rm cm}^{-2}$.  It is an eclipsing
X-ray pulsar ($P_{{\rm spin}}=139.612$~s) which revolves
the donor star each 4.6 d \citep{hill05}. 

Based on the orbital solution, \citet{hill05} proposed that
the mass donor should be an early type supergiant with most likely
parameters $M_{*}\sim21\,M_{\sun}$ and $R_{*}\sim19\,R_{*}$. \citet{masetti08}
identified the counterpart as 
2MASS~J18024194$-$2017172 and obtained a low-resolution optical spectrum
that showed it to be a very reddened early-type star. Using the mass
and radius deduced by 
\citet{hill05}, they argue that the counterpart should be a B-type
giant. \citet{chaty08} confirmed the identity of the 
counterpart.  Their infrared spectrum seems typical of an
early-B supergiant, displaying prominent Brackett lines in the $H$
band even though these authors claimed the detection of \ion{He}{i} and
\ion{He}{ii} lines in the optical spectrum. 
In fact, they favor a B type supergiant based on a $T_{{\rm eff}}\sim20800$~K deduced
from a fit to the SED. Based on the $R_{*}/D$ value of the normalization,
and assuming a stellar radius $R_{*}=20\,R_{\sun}$, they estimate a distance
of $\sim 12$~kpc. \citet{chaty08} estimate $A_{V}\sim 8.8$~mag, while
\citet{masetti08} estimate $\sim 8.3$~mag.  In the following, we
will use our TNG $K$-band and WHT $I$-band spectra
(Fig.~\ref{fig:18027_both_spec}) to refine
the classification of the counterpart and establish the nature of the system.   

\textsc{$K$ band spectrum}. The presence of \ion{He}{i} lines and the absence
of \ion{He}{ii} (which is seen up to O9) points towards a B type
star. The Br$\gamma$ line is in emission. This line seems to be
blended with the blueward absorption of the \ion{He}{i}~2.161$\mu$m
line. This morphology is more pronounced 
in high luminosity B stars
\citep[][Fig.~12]{hanson05}. \ion{He}{i}~2.058$\mu$m is also in
emission, with a possible P-Cygni profile, a feature also noted by
\citet{chaty08} on their lower resolution spectrum. In the atlas of 
\citet{hanson05}, no luminosity class III star shows
\ion{He}{i}~2.058$\mu$m in emission. In contrast, this line is in
emission in B supergiants and Be
stars. However, \ion{He}{i}~2.113$\mu$m is not seen in Be stars, while it is
very prominent in our object, a feature of B supergiants \citep[see
  Figures 13 to 16 in][]{hanson05}. We therefore conclude
that the counterpart is an early B supergiant star.

\textsc{$I$ band spectrum}.  As seen in Fig.~\ref{fig:18027_both_spec}, our
$I$-band spectrum has a low SNR. However, it is easy to see the narrow
well-defined Paschen lines, typical of a high-luminosity object. Moreover,
Paschen lines are visible up to Pa~19, in spite of moderate SNR and
resolution, clearly identifying the star as a supergiant (compare to
the B1\,V star in the same figure). The prominent \ion{He}{i} lines
are typical of supergiants with spectral types close to B2 (compare to
the earlier supergiants in the same figure). Moreover, as can be seen
in the spectra of the O9.5\,Ia, for stars B0.5 and
earlier, the \ion{C}{iii}~8502\AA\
is blended with Pa~16, making it stronger than the
neighbouring Pa~15 and Pa~17. This is not the case in our object,
suggesting it is later than B0.5.  On the other hand, the red spectrum
of IGR~J18027$-$2017 does not display \ion{O}{I}~7774\AA\ (which is
not seen in the spectrum of \citet{chaty08} either). This would indicate a
type earlier than B1.5 if it is a luminous supergiant or earlier than B2 if it is less luminous.


Taking together all the data, we propose a B1 Iab - B1 Ib supergiant
counterpart. To discriminate between the two luminosities is very difficult with the available spectra. However, the eclipse duration covers $\sim0.25$ of the orbit \citep{hill05}. With a period of only 4.6 d, a B1 Iab supergiant would not fit into the orbit. Therefore, a luminosity class Ib is definitely assigned to the donor.
This locates the source at a 
distance of $d\approx 12.4$ kpc, in agreement with \citet{chaty08}, between 
the Sagittarius and Perseus arms, well beyond the Galactic Center (Fig. \ref{fig:galaxy}).

Given this classification, the observed
  color excess is $E(B-V)\approx 3$ and, assuming a standard extinction law,
the extinction in the visual band would be $A_{V}\approx 9.4$. This is
slightly larger than the value given by \citet{chaty08} ($A_{V}\approx 8.8$) and by \citet{masetti08}, namely $A_{V}\approx 8$. 

For a standard reddening law \citep{fitzpatrick}, and typical
intrinsic colours $(I-J)_{0}=-0.15$, $(R-I)_{0}=-0.17$ \citep{ducati01},
our value $E(J-K)\approx1.54$ implies $I\approx14.8$ and $R\approx16.9$ in excellent agreement with the value $R=16.9\pm0.1$ measured by \citet{masetti08}.

\subsection{SAX~J18186$-$1703}

SAX~J1818.6$−$1703 was discovered by {\it BeppoSAX} during a strong
outburst with a fast rise time of $\sim$ 1 h, in March 1998
\citep{intzand98}.  Subsequently {\it INTEGRAL} detected a double-peaked 
outburst in September 2003 \citep{gs05} and 
two more in October 2003 \citep{sgue05}. Other fast outbursts have been observed with the ASM onboard {\it RossiXTE}
\citep{sgue05}. Its counterpart, 2MASS~J18183790$-$1702479, was identified by a
{\it Chandra} localization \citep{intzand06a}. Recently \citet{zurita09}, using {\it INTEGRAL}, have reported a 22--50\,keV flux of the order of (2--8)$\times 10^{-11}\:{\rm erg}\,{\rm s}^{-1}\,{\rm cms}^{-2}$ in quiescence while it reaches (1--2)$\times 10^{-9}\:{\rm erg}\,{\rm s}^{-1}\,{\rm cms}^{-2}$ in the strongest flares. No pulsations have been detected so far, but an orbital period of 30 d has been established \citep{zurita09}.

\begin{figure}
   \centering
   \includegraphics[width=\columnwidth]{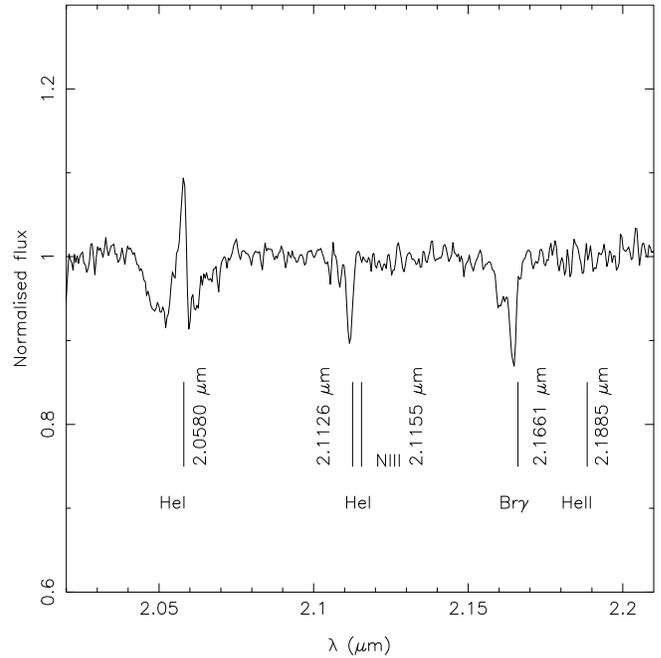}
   \caption{$K$ band spectrum of the counterpart to SAX~J18186$-$1703
     with the most relevant features expected in this range
     identified. Note the absence of any feature at the position of
     the He \textsc{ii} 2.1885$\mu$m line.} 
    \label{fig:saxj1818_k}
\end{figure}  

\textsc{$K$ band spectrum}. In Fig.~\ref{fig:saxj1818_k}, we present
the $K$ band spectrum of the IR counterpart. The absence of
\ion{He}{ii}~2.1885$\mu$m confirms that this star is later than $\sim$ 
O8. 
The He \textsc{i} 2.0581$\mu$m line is strongly in
emission. This is typical of early B supergiants. Along with the
morphology of the complex He \textsc{i} 2.1607/2.1617 - Br$\gamma$ it
points to high luminosity. Main sequence early B stars have broader
lines \citep[Fig.~12]{hanson05}. Following the same arguments
presented for IGR~J18027$-$2017, we assign a similar spectral type to
this object and conclude it is a supergiant in the B0--B1.5 range. The
exact luminosity class is difficult to determine, but, as discussed previously,
it is likely to be at least Iab. This would place the source at a distance of $d\approx 2.1$ kpc in the outskirts of the Sagittarius arm (Fig. \ref{fig:galaxy}).

\subsection{IGR~J18483$-$0311}

IGR~J18483$-$0311 is a transient X-ray source discovered by
\citet{chernya03} during observations of the
Galactic Center with {\it INTEGRAL}. Using \emph{Swift}, \citet{sguera07} suggest as a counterpart the IR source
2MASS~J18481720$-$0310168,  which is confirmed by \citet{chaty08}. The
X-ray source shows very high obscuration 
($N_{\rm H}=9^{+5}_{-4}\times 10^{22}\:{\rm cm}^{-2}$). The system contains a pulsar 
($P_{{\rm s}}=21.0526\:{\rm s}$) in orbit around the primary with a period of $P_{\rm orb}=18.52 d$
\citep{sguera07}. The unabsorbed X-ray 20--100~keV flux, measured with
IBIS on board {\it INTEGRAL}, is $\sim 2\times 
10^{-9}\:{\rm erg}\;{\rm s}^{-1}\;{\rm cm}^{-2}$ \citep{sguera07}.

\begin{figure}
   \centering
   \includegraphics[width=\columnwidth]{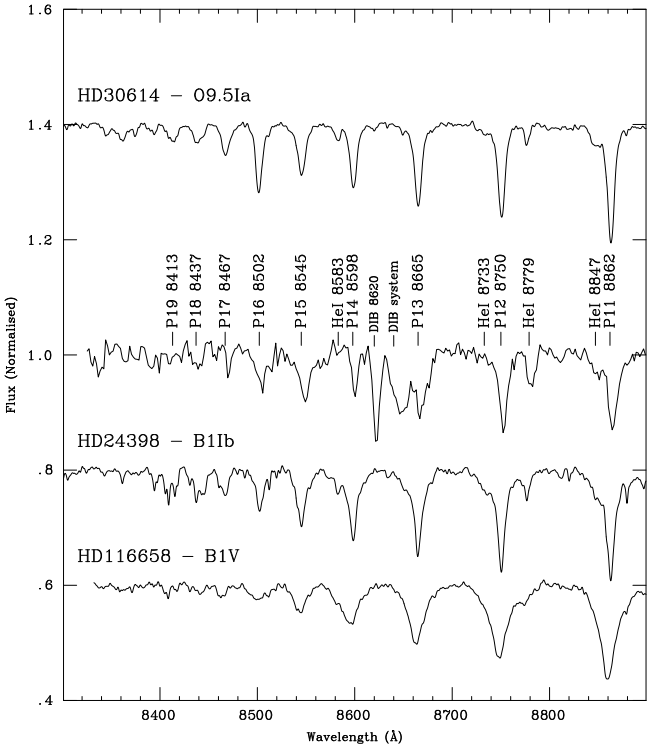}
   \caption{$I$ band spectrum of IGR~J18483$-$0311, compared to
     several other objects with well-determined spectral types. Note
     the similarity of the spectrum to the B1\,Ib star. Note also the
     strong DIBs, indicating a heavy interstellar absorption.}
    \label{fig:igrj1848_I}
\end{figure}

By comparing a low-resolution $HK$ spectrum to spectra of
classification standards, \citet{rahoui08} conclude
that the counterpart is a B0.5 supergiant, and suggest a luminosity
class Ia.  In Fig.~\ref{fig:igrj1848_I}, we present the $I$ band
spectrum of this object, bracketed between those of two supergiants
with well-determined spectral type. Our spectrum shows narrow
well-defined Paschen lines, typical of a high-luminosity object. Moreover,
Paschen lines are visible up to Pa~18, in spite of moderate SNR and
resolution, clearly identifying the star as a supergiant (compare to
the B1\,V star in the same figure). The presence of strong
\ion{He}{i}~8779\AA\ is typical of early B supergiants, while the
absence of \ion{O}{i}~8446\AA\ and \ion{N}{i} lines makes it earlier
than B2. Pa~16 is not prominently stronger than Pa~17, suggesting that
it does not contain an important contribution from
\ion{C}{iii}~8502\AA. As discussed above, this suggests
a spectral type B0.5 or later. The exact determination of the luminosity
class is not easy with the data available. However, the fact that we
only detect Paschen lines up to P18 suggests that the star is not a Ia
supergiant. Combining our data with the $HK$ spectrum of \citet{rahoui08}, we conclude that the star is a B0.5--B1
supergiant, most likely with a luminosity class Iab. This would locate the system at $d\sim$ 2.8 kpc.

\subsection{IGR~J1914.0+0951}

\begin{figure*}
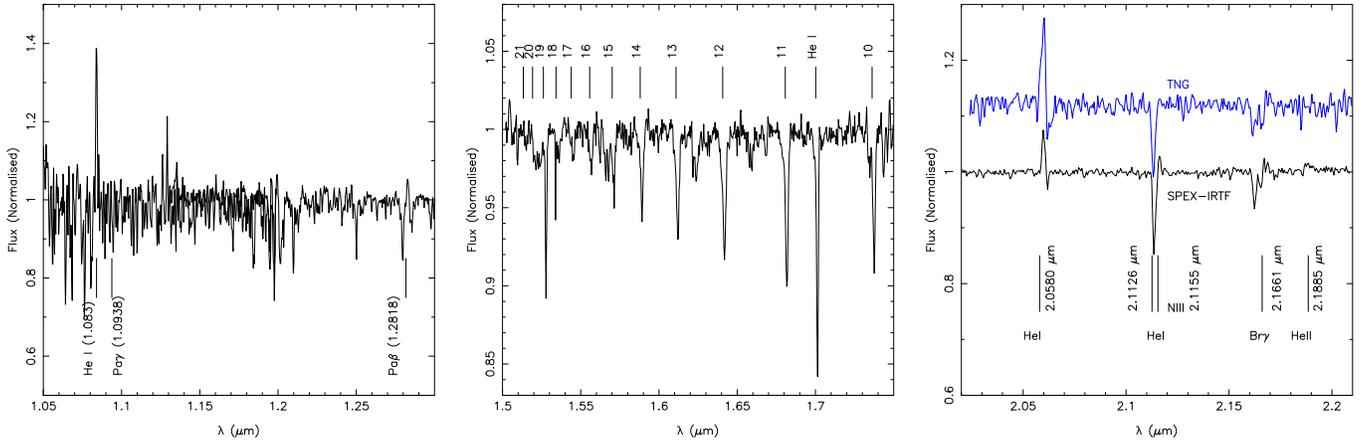

   \centering
   \includegraphics[width=0.67\columnwidth,clip]{19140_J.ps}
   \includegraphics[width=0.67\columnwidth,clip]{19140_H.ps}
   \includegraphics[width=0.67\columnwidth,clip]{19140_K.ps}
   \caption{SPEX-IRTF $J$, $H$ and $K$ band spectra of IGR~J19140$+$0951. In the
     $K$ band, TNG data (top) are also presented. } 
    \label{fig:19140_spec}
\end{figure*}

IGR~J19140+0951 was discovered by \citet{hannika04}
with {\it INTEGRAL}. It displays X-ray characteristics typical of a
SGXB. \citet{intzand06b} used \emph{Chandra} to
pinpoint the IR counterpart, which turns out to be
2MASS~J19140422+0952577. A period of 13.55 d is observed with {\it
  RossiXTE} \citep{corbet04}. No pulsations have been
detected so far. 

\citet{intzand06b} presented a low resolution red spectrum of the source,
that was very badly affected by fringing. It shows no counts bluewards
of 7000\AA. They also quote a magnitude $I=13.0$ for the counterpart,
completely at odds with our value, and the low counts in the
corresponding region of the spectrum. 

Unfortunately, our $I$-band spectrum, though not affected by fringing
is of low resolution and SNR. We identify a few narrow, strong Paschen lines,
that indicate  an early-type supergiant. We do not detect the 
\ion{O}{i}~7774\AA\ line, which (given the low SNR in this part of the
spectrum) constraints it to be earlier than $\sim$B3. 


\citet{hannikainen} have presented a low SNR $K$-band spectrum and an
$H$-band spectrum of much better resolution and SNR. Their analysis
led them to conclude that the star was a B0.5\,Ia supergiant.
\citet{nespoli08} classify it as B1\,Iab, based on a $K$-band
spectrum. The spectrum from \citet{hannikainen} displays 
Br$\gamma$ weakly in absorption, while this line is more clearly in
absorption in the spectrum of \citet{nespoli08}.
 However, our TNG spectra,
taken four months later, as well as our SPEX spectra, taken one
year earlier (both of higher SNR), show a P-Cygni profile in this line. This is
indication of high activity in this object and further supports the 
high luminosity of the star. 


No \ion{He}{ii} absorption lines are seen in our spectra, suggesting
that the star is later than O9. The presence of a very well defined
series of Brackett \ion{H}{i} lines in the $H$ band (see
Fig. \ref{fig:19140_spec}, $H$ band panel) also points towards an early B star, as
these lines disappear for O type stars. The narrowness and depth of
the Br lines is a clear indication of high luminosity. Moreover, we
detect them well up to Br19, at least. The narrow and deep
\ion{He}{i}~1.7$\mu$m line is also typical of a supergiant.  It is deeper
than neighbouring Br lines, indicating high luminosity.  Its relative depth with respect nearby Br 11 points clearly towards type B0-B0.5Ia and rules out types earlier than 09.7I \citep{hanson98} (Fig. 2).

In addition, we find \ion{He}{i}~2.0581$\mu$m strongly in emission, a
characteristic of luminous early-B supergiants. This emission seems to be variable, as it is much more pronounced in the TNG data than in the IRTF observations (Fig.\ref{fig:19140_spec}). We also see emission
in the \ion{He}{i}~1.083$\mu$m line in the $J$ band, and in
Pa$\beta$. 

\begin{figure*}
   \centering
   \includegraphics[width=\columnwidth,clip]{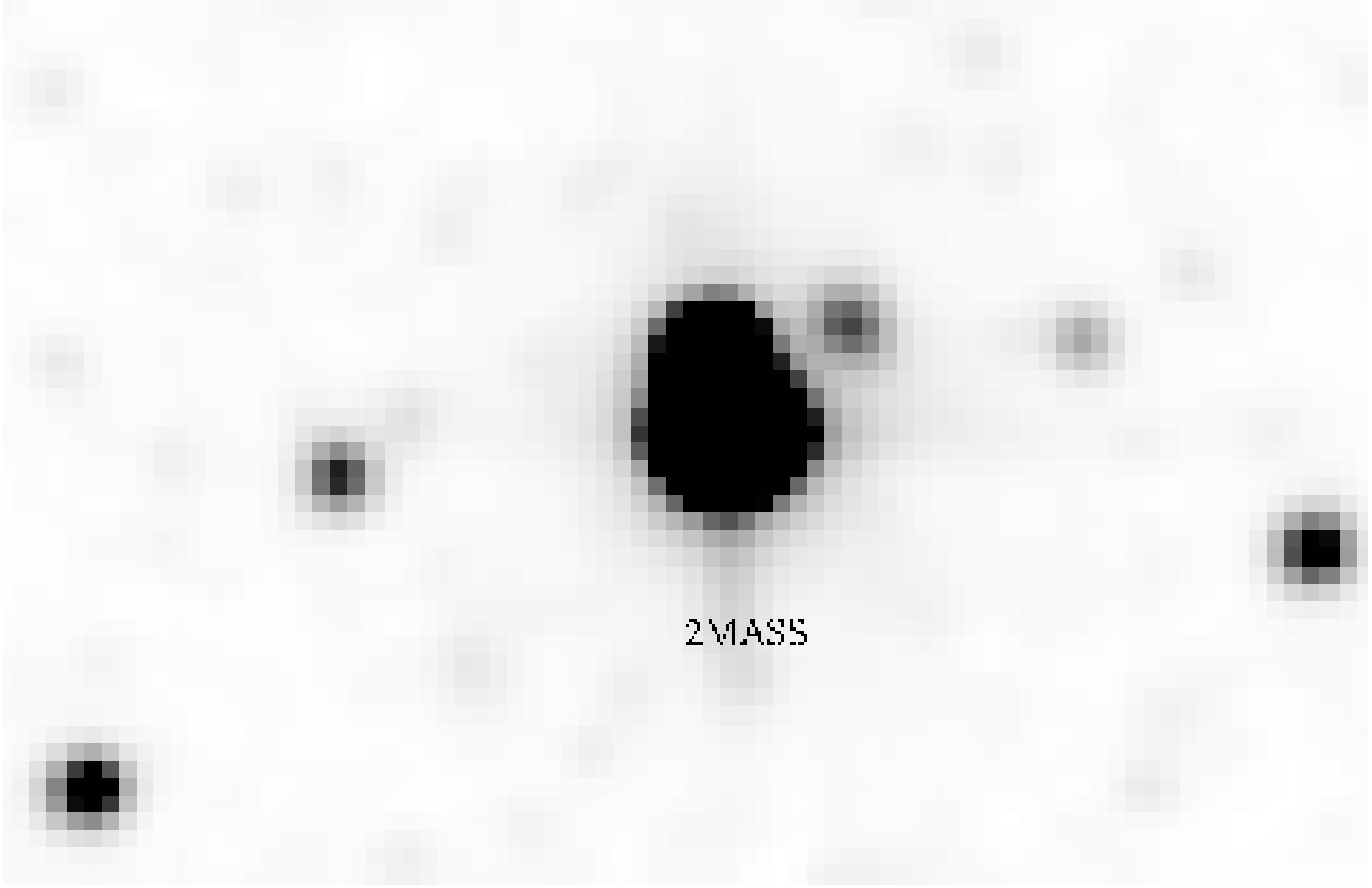}
   \includegraphics[width=\columnwidth,clip]{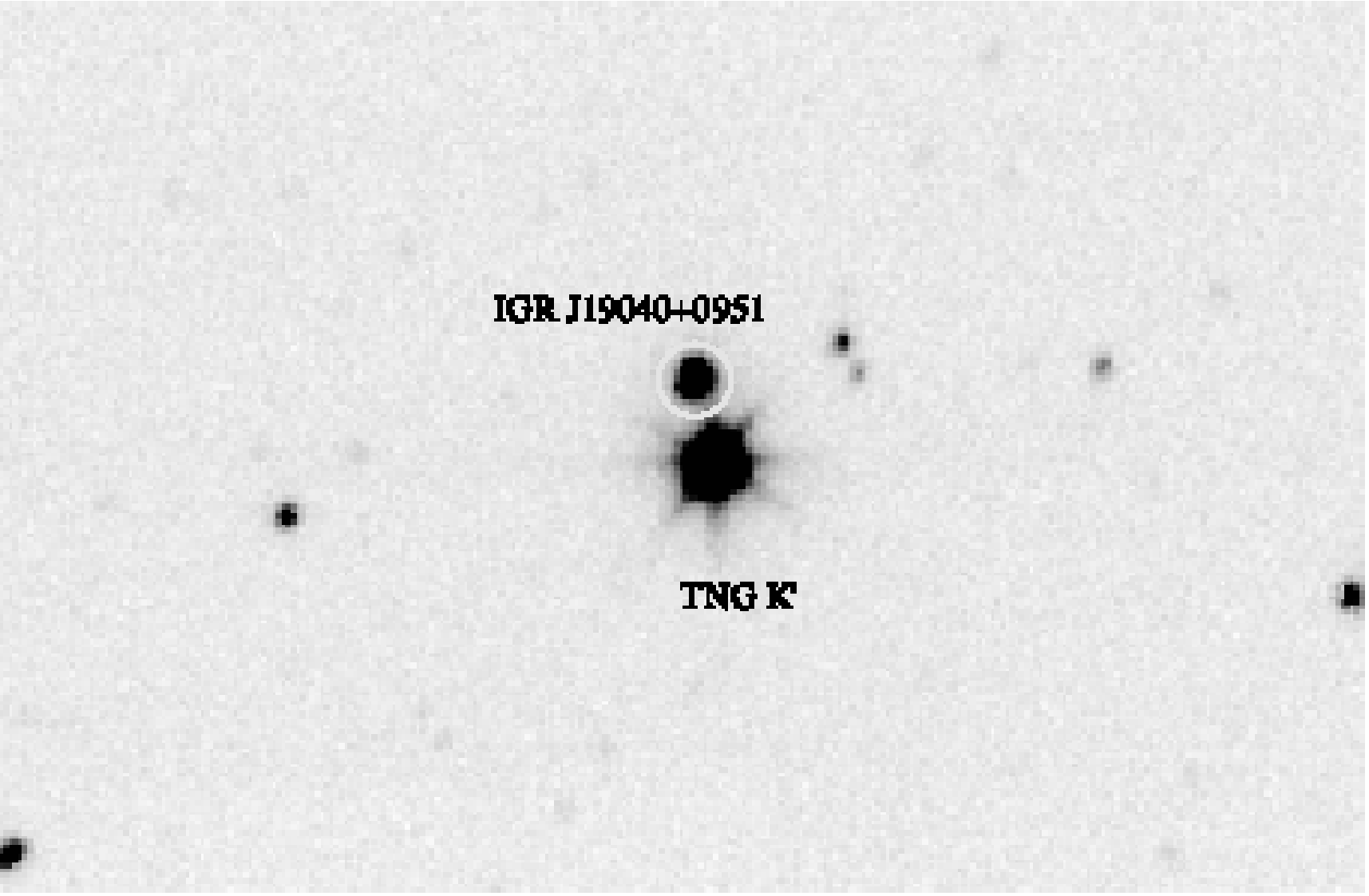}
   \caption{The counterpart to IGR~J19140+0951. {\bf Left panel:
     }$K$-band image from 2MASS, showing the contamination by the
     light of the bright star 2MASS~J19140417+0952538, which affects
     the catalogue photometry for this source. {\bf Right panel: }TNG
     $K^{\prime}$ band image, showing the two stars clearly
     resolved (cf. Fig.~2 in \citet{rahouietal08}.)}
    \label{fig:19140_finder}
\end{figure*}

We conclude that the spectral type is B0.5\,Ia, compatible with that
given by \cite{hannika04} and \citet{nespoli08}, but
with a slightly higher luminosity class. \citet{nespoli08} estimate a
distance of $\approx 1.1$ kpc, but they use the 2MASS photometry,
contaminated by the very 
bright star towards the upper right (see Fig. \ref{fig:19140_finder}),
namely 2MASS~J19140417+0952538 ($K_{S}\approx 6.5$ mag), as was
already recognised by \citet{intzand06b}. 
We performed our photometry on images of
higher spatial resolution which allowed us to clearly resolve both
stars. We used an aperture small enough to ensure that the
contribution from the bright star is minimal. Then we applied aperture
correction to our magnitudes keeping in mind that avoiding completely
the contamination is probably impossible without adaptive optics.
Using our photometry, rather than the 2MASS values, and the deduced spectral type, 
we compute a distance of $d\approx 3.6$~kpc.

\begin{figure}
   \centering
   \includegraphics[width=\columnwidth,clip]{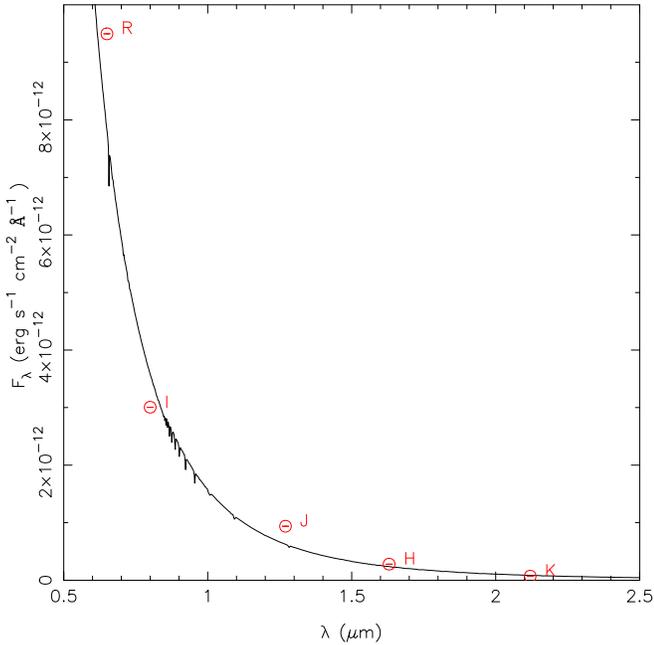}
   \caption{Our $RIJHK$ photometry of IGR~J19140$+$0951 
     dereddened with a standard extinction law and $E(B-V)=5.5$ appears
     superimposed on a Kurucz
     atmosphere model for $T_{{\rm eff}}=28000$~K and $\log g=3.5$,
     corresponding to an early type B supergiant. } 
    \label{fig:19140_fotom}
\end{figure}

In Fig. \ref{fig:19140_fotom} we make use of our red and infrared
photometry for this object to plot the $RIJHK$
photometry absolute flux calibration, dereddened with a standard
reddening law and taking $E(B-V)=5.5$. The photometry is
superimposed on a Kurucz atmosphere model for $T_{eff}=28000\:{\rm K}$ and
$\log g=3.5$, corresponding to an early type B supergiant\footnote{Actually, $\log g=3.0$ would be more adequate. Unfortunately, this Kurucz model is not available. The difference in slope, however, is negligible for the present purposes}. The acceptable fit shows that the extinction to this
source is close to standard, in spite of the high value. The
normalization constant in Fig.~\ref{fig:19140_fotom}, assuming 
$R_{*}\approx 30\,R_{\sun}$, would yield a distance of $\approx
3.6$~kpc, in agreement with the value computed previously. The
independent estimate by \citet{rahouietal08}, based on a fit to the near- and
mid-infrared magnitudes, of $d\sim3.1$~kpc, is also consistent with our
value.

\subsection{XTE~J1901$+$014}

\begin{figure}[ht]
   \centering
   \includegraphics[angle=0,width=\columnwidth]{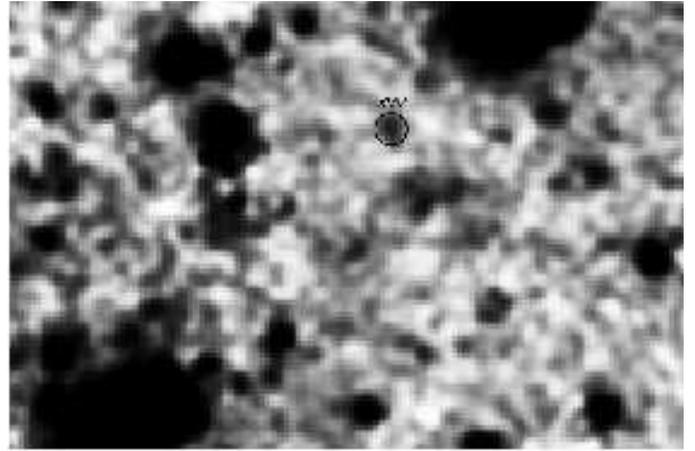}
   \caption{TNG $K^{\prime}$-band image of the field around
     XTE~J1901+014. The blue circle
     represents the nominal {\it XXM-Newton} position with an error
     radius of $1\arcsec$ (which for our image scale is 4 pixels). A
     faint possible candidate is seen inside the {\it XMM-Newton}
     error circle. The candidate must have $K^{\prime}\gtrsim 14.2$. It is not visible in either $J$ or $H$ bands. }
              \label{fig:xte1901_pos_cand}
    \end{figure}

The source XTE~J1901+014 was discovered by the {\it RXTE}/ASM monitor
in 2002 during a brief outburst, after which the source went into
quiescence below detectable limits.  Examination of archival data
showed that the source had undergone a previous brief outburst in 1997
\citep{rs02} showing a behaviour reminiscent of that of 
SFXTs. \citet{smith07}) used {\it XMM-Newton} to obtain
accurate position of the X-ray source, which turns out to be
$\alpha=19\:{\rm h}$~$01\:{\rm m}$~$40.20\:{\rm s}$, $\delta=
+01\degr\:26\arcmin\:26\farcs0$ 
with an error of $\sim 1\arcsec$. Subsequently we searched the area for possible
counterparts. In Fig. \ref{fig:xte1901_pos_cand}, we show a $K^{\prime}$-band
image of the area encompassing the error circle. Only one possible
candidate is visible within the {\it XMM-Newton} error circle. The
object is very faint with $K^{\prime}\gtrsim 14.2$, close to the limit of the
sky brightness at 2$\mu$m at the La Palma Observatory. No
sources show up in the $J$- or $H$-band images.  

As the infrared observations do not provide enough information, in
order to guess the nature of the system we will resort to X-ray
data. The {\it XMM-Newton} spectra can be well fitted by a thermal
Bremsstrahlung with $kT \sim 6$~keV and an absorption column of
$N_{{\rm H}}\sim 2.5\times 10^{22}\:{\rm cm}^{-2}$ (Rampy 2009,
priv. comm.). This would correspond to  
$E(B-V)\approx 3.7$ \citep{ryter96}, which is larger than the
integrated value for the Milky Way Galaxy in the direction of the
source, namely $\sim 8.5\times 10^{21}\:{\rm cm}^{-2}$ suggesting intrinsic
absorption, as might be expected. Assuming that all absorption is interstellar, the corresponding infrared excess is
$E(J-K)\approx 1.8$. Comparing these values with those shown in Table
\ref{tab:data}, we can see that the source is, indeed, very reddened but not
particularly so. 

If we assume an$M_{K}$ typical for a not very luminous OB supergiant \citep[$\sim
-5.5$;][]{martins06}, with $K\approx
14.2$, the distance to the source would be $\sim 47$~kpc, placing it
well outside the Galaxy. This seems very unlikely. At the outskirts of
the Galaxy in the direction to the source ($\sim 22$~kpc), the infrared
magnitude of the counterpart would be $M_{K}\approx -3.8$. This would
correspond to a O8\,V star or a K5\,III star. Therefore the
possibility of a supergiant companion is ruled out.

However, there are arguments to believe that the distance must be much
lower. Indeed, the 2002 outburst detected with {\it RossiXTE}/ASM reached
a luminosity of
the order of 1 Crab, which corresponds to a flux $F_{X}=3.63\times
10^{-8}\:{\rm erg}\,{\rm cm}^{-2}\,{\rm s}^{-1}$. At a distance
$d=22$~kpc, the outburst should have had an $L_{X}\approx 2\times
10^{39}\:{\rm erg}\,{\rm s}^{-1}$. This
is equivalent to the Eddington luminosity for a compact object of $\sim
16M_{\sun}$, and higher than ever reported for any neutron star
system. On the 
other hand, if we assume that the outburst luminosity was at the
Eddington luminosity for accretion onto a NS ($\sim 2\times
10^{38}\:{\rm erg}\,{\rm s}^{-1}$), the maximum distance to the source
should be of the 
order of $d_{{\rm max}}\sim 7$ kpc, which would place the source at the tip
of the Long Bar in the galactic nucleus. For such a source,
$M_{K}\approx -1.28$ which would be compatible with a $\sim$G5\,III star. 
In fact, the maximum distance at which the primary would be compatible with a class III star is  $d\sim 5$~kpc, in which case, we would have G4III (or also B4V). Below this distance, essentialy, only main sequence stars are allowed. For distances around 1 kpc or less, the only compatible companions would be G, K or M main sequence stars. This raises the interesting possibility that the system is, actually, a Cataclysmic Variable, since the X-ray spectrum can be described by a thermal Bremsstrahlung at 6 keV. However, in such a case, the $L_{\rm X}\sim 10^{36}\:{\rm erg}\,{\rm s}^{-1}$ which would be several orders of magnitude larger than that usually found in CV.  Therefore, we conclude that, very likely, $d_{{\rm
    min}}\sim 1-2$~kpc. This is in good agreement with the conclusions obtained by \citet{karasev08}, who argue in favor of a distance of $\sim 5$~kpc.

\begin{figure}[ht]
   \centering
   \includegraphics[angle=0,width=\columnwidth]{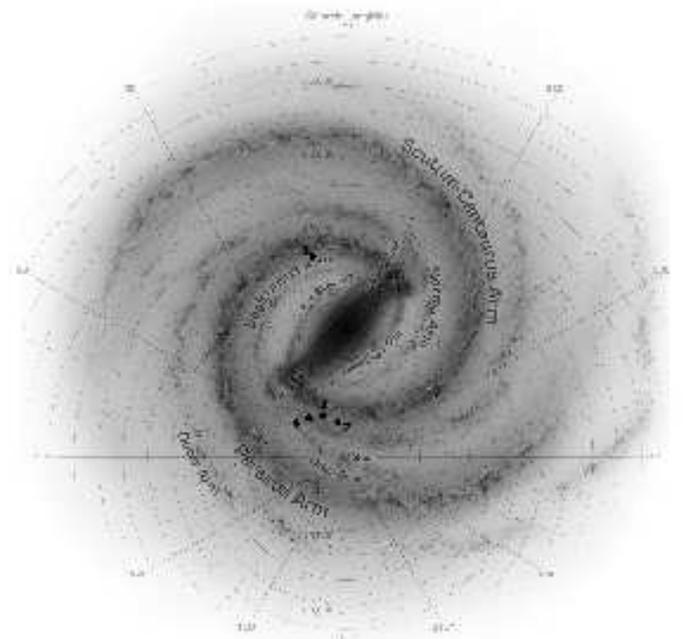}
   \caption{Sketch of the Galaxy showing the position of the HMXBs studied in this work (filled dots). Numbers refer to the relative order of the source in Table \ref{tab:data}. Galaxy model from \citet{church09}.  }
              \label{fig:galaxy}
    \end{figure}

\section{Discussion and conclusions}

    \begin{table*}
      \caption[]{Derived parameters for stars in our sample,
        calculated using the observed data.  }
         \label{tab:data}
 	\begin{center}   
         \begin{tabular}{llccccccrr}
            \hline
	    \hline
            \noalign{\smallskip}
 Source & Spectral type & Class & $E(B-V)$ & $E(J-K)$ & $M_{V}$ & $M_{K}$ & $d$ (kpc) & $L_{X}$ (erg/s)$^{\rm (a)}$ \\
            \noalign{\smallskip}
            \hline
            \noalign{\smallskip}
 IGR~J18027$-$2016 & B1\,Ib & SGXB & 3.04$\pm0.02$ & 1.54 & $-5.8$ & $-5.1$  & 12.4$\pm 0.1$ &  (1.65$\pm0.03$) $\times 10^{36}$ \\
 SAX~J18186$-$1703 & B0.5\,Iab & SFXT & 5.08$\pm0.05$ & 2.54 & $-6.4$ & $-5.65$ & 2.1$\pm0.1$ & (8$\pm0.4$) $\times 10^{35}$ \\
 IGR~J18483$-$0311 & B0-1\,Iab & SFXT & 5.22$\pm0.02$& 2.61 & $-6.4$ & $-5.66$ & 2.83$\pm 0.05$ & (1.9$\pm0.06$) $\times 10^{36}$ \\
 IGR~J19140$+$0951 & B0.5\,Iab-a & SGXB & 5.5$\pm0.1$ & 2.75 & -6.65$\pm 0.03$ &  -5.90$\pm0.25$ & 3.6$\pm 0.04$ & (1.6$\pm0.3$) $\times 10^{36}$ \\
\hline
\noalign{\smallskip}
XTE~1901$+$014 & G5\,III - A3\,V & LMXB? & $\sim$3.7 & 1.8 &  &   & 2-7 & 1$^{+1}_{-0.8}\times 10^{38}$  \\ 
            \noalign{\smallskip}
            \hline
         \end{tabular}
\begin{list}{}{}
\item[$^{\mathrm{a}}$] Observed peak luminositites for the energy ranges (2--10)\,keV, (22--55)\,keV,  (20--100)\,keV, (2--20)\,keV and (3--12) \,keV respectively.

\end{list}
        \end{center}
   \end{table*}

Four of our sources have early-B supergiant
companions. 
All four counterparts turn out to cover a rather narrow spectral
interval, being early-B (B0--B1.5) supergiants of moderate or high
luminosity. This identifies the nature of the X-ray source
unambiguously as HMXBs. Furthermore, they are all heavily obscured,
with $E(B-V)\sim 3-5.5$, implying extinctions on the order of $A_{V}=9-16$
mag in the visual band. But, on the other hand, XTE~1901+014 has no obvious counterpart in any band
except $K^{\prime}$. Thus, a supergiant nature is definitely excluded, thereby setting it apart from the other four. 
  

IGR~J18027$-$2017 is a persistent X-ray source with a counterpart
around B1\,Ib. It is therefore, a SGXB system. 
 The reason why it has
not been detected by earlier X-ray missions is not clear. For example,
it was not detected by \emph{ROSAT}. At a 
distance of $d\approx 12.4$ kpc, its {\it XMM-Newton}
detected flux in the 2--10~keV band for the main pulse of $8.9\times
10^{-11}\:{\rm erg}\,{\rm s}^{-1}\,{\rm cm}^{-2}$, would yield a luminosity of
$L_{X}=1.65\times 10^{36}\:{\rm erg}\,{\rm s}^{-1}$. This is typical for
these kind of systems albeit in the lower end. Since the detected flux
outside the primary pulse is a little bit smaller, the system can
spend a fraction of its time in the upper $\times 10^{35}$ erg
s$^{-1}$. Given its high obscuration, this can explain why it was
missed by earlier surveys.


For SAX~J18186$-$1703, the NIR counterpart is a B0.5Iab star, confirming its nature as a SFXT. The 22--50\,keV flux reported by \citet{zurita09} is of the order of (2--8)$\times 10^{-11}\:{\rm erg}\,{\rm s}^{-1}\,{\rm cms}^{-2}$ in quiescence while it reaches (1--2)$\times 10^{-9}\:{\rm erg}\,{\rm s}^{-1}\,{\rm cms}^{-2}$ at the strongest flares. At the deduced distance of $\sim 2.1$ kpc, the X-ray luminosity of this
object would be $\sim 3\times 10^{34}\:{\rm erg}\,{\rm s}^{-1}$ in quiescence, while the peak luminosity can be as high as $\sim 8\times 10^{35}\:{\rm erg}\,{\rm s}^{-1}$. This is lower than those found in other SFXTs ($\sim 10^{36}\:{\rm erg}\,{\rm s}^{-1}$). \citet{zurita09} have found that this system is in a very eccentric orbit ($e\sim 0.3-0.4$), with $P_{orb}\approx 30$ d, amongst the largest yet found for SFXT, reaching a periastron distance between 2 and 3 $R_{*}$. This would
locate the compact object at a slightly larger distance from the donor than in the SGXB systems ($a\lesssim 2R_{*}$) thereby reducing the $L_{X}$ of the outbursts slightly below $10^{36}\:{\rm erg}\,{\rm s}^{-1}$ as is observed.


IGR~J18483$-$0311 is a transient system with a B0--B1\,Iab primary. It has been classified as an intermediate SFXT \citep{rahoui08}. Like many other
SFXTs, it is a nearby source at $\sim$ 2.8 kpc. For this distance, the
{\it INTEGRAL}/IBIS 20--100~keV flux \citep{sguera07} translates into a luminosity  $\sim 1.9 \times
10^{36}\:{\rm erg}\,{\rm s}^{-1}$, for the strongest flares, which is typical for this kind of
objects. The orbital period of this system is 18.5 d, while the spectral type of the companion is similar to that of SAX~J18186$-$1703. Therefore, the compact object would be located slighly closer to the star and, correspondingly, its average X-ray luminosity will be higher, in agreement with the scenario described in \citet{neg08}.


IGR~J19140$+$0951 is a persistent X-ray source and therefore, owing to
its counterpart, a SGXB located at $d\approx 3-4$~kpc, a
little bit further than previously thought. For the bright {\it
  INTEGRAL} flux in the 2-2-0~keV band of $1\times
10^{-9}\:{\rm erg}\,{\rm s}^{-1}\,{\rm cms}^{-2}$ \citep{hannika04},
the $L_{X}=1-2\times 10^{36}\:{\rm erg}\,{\rm s}^{-1}$, typical of a
    SGXB albeit, again, in the lower end.  The fact that, together with IGR 18027-2017, they
   have lower X-ray fluxes than the average SGXBs explains their
   more recent detections in the newer, and deeper {\it INTEGRAL} observations.

 As can be seen in Fig.~\ref{fig:galaxy}, the sources tend to concentrate in or behind the section of the Sagittarius arm projected onto the direction of the Galactic Bulge, an area intensively scanned by {\it INTEGRAL}. They are, thus, relatively close to the Sun (in comparison to typical
          distances to HMXBs), although highly reddened. 
This fact can introduce a
          bias in the newly detected populations of obscured
          sources, which tend to harbor SG companions, as only the
          brighter and/or closer ones can be reached with telescopes
          of moderate apertures in the 4~m class. A transient system with a main sequence companion, like a BeXB, will be too faint to be conclusively classified \citep[i.e.][]{zurita08}. Remarkably, the furthest source in our sample is the least reddened one. This raises the question of the true abundance of HMXBs (and OB stars in fact) in our Galaxy. As can be seen in Table \ref{tab:data}, the high reddening is not a sufficient condition to be an SFXT since some SGXBs (like IGR~J19140$+$0951) are more reddened. On the other hand, it is not a \emph {necessary} condition either, as other SFXTs present rather low reddenings \citep[for example IGR~J08408$-$4503 with $E(B-V)=0.5$][]{neg09}. Furthermore, the position in the arms of the Galaxy of the newly discovered HMXBs (including the SFXTs) is entirely consistent with the positions of the previously known HMXBs (including SGXBs). Therefore there is no reason to believe that they form a different population. This would be expected from a scenario like that discussed in \citet{neg08}. In this scenario, the SFXTs are explained as just SGXBs where the compact object is located further out in the system, in a region where the stellar wind of the supergiant companion has become substantially clumped. The number of HMXBs in the Galaxy would then be much larger than previously thought. But so would the number of isolated OB stars whose much lower X-ray emission will not show up in the current X-ray surveys if they are heavily obscured. This is supported by the discovery, in the last few years, of a growing population of very massive infrared clusters \citep[e.g.][]{crow06, mess08}.  Therefore, in principle, there is no reason to believe that the relative abundance of HMXBs, with respect to OB stars, differs from the current values which are well explained by the theoretical models.


Finally, XTE~J1901$+$014 emerges as a transient source whose X-ray
behaviour is characteristic of the SFXTs but where the presence of a
Supergiant companion is clearly ruled out. It can still be a massive
star close to the main sequence in
the outskirts of the Galaxy, perhaps with a $16\,M_{\sun}$ BH as a
companion. In this respect, it is interesting to note the case of the LMC transient A\,0535$-$66
reaching  $L_{X}\approx 8\times 10^{38}\:{\rm erg}\,{\rm s}^{-1}$
\citep[e.g.,][]{pp81}. A similar source at the edge of the Galaxy
could have a similar hard X-ray behaviour. However, taking into
account the moderate hardness (the source is seen as 1RXH~J190140.1+012630) and obscuration of the source, it
seems more likely that the counterpart is a late type star
located between 2~kpc and the Long Galactic bar (7~kpc), somewhere between luminosity class V to class III, but definitely not consistent with a class I
star. Whether a main sequence or giant late type star wind can produce
the instabilities necessary to produce the observed outbursts via
accretion onto a compact object remains an issue.

\begin{acknowledgements}
This research has been funded by grants AYA2008-06166-C03-03 and
Consolider-GTC CSD-2006-00070 from the Spanish Ministerio de Ciencia e
Innovaci\'on (MICINN). JMT also acknowledges the research grant PR2007-0176. The authors would like to thank the referee, Dr. Alex Kaper, for his useful comments and suggestions.

The Nordic Optical Telescope is operated 
on the island of La Palma jointly by Denmark, Finland, Iceland,
Norway, and Sweden, in the Spanish Observatorio del Roque de los
Muchachos of the Instituto de Astrofisica de Canarias. The data were
taken with ALFOSC, which is 
owned by the Instituto de Astrof\'{\i}sica de Andaluc\'{\i}a (IAA) and
operated at the Nordic Optical Telescope under agreement
between IAA and the NBIfAFG of the Astronomical Observatory of
Copenhagen.

The WHT is operated on the island of La 
Palma by the Isaac Newton Group in the Spanish Observatorio
del Roque de Los Muchachos of the Instituto de
Astrof\'{\i}sica de Canarias. The WHT data presented here have been
obtained as part of the service programme. We thank the support
astronomers for their dedication during the service nights.

Partly based on observations made with the Italian Telescopio Nazionale
Galileo (TNG) operated on the island of La Palma by the Fundaci\'on
Galileo Galilei of the INAF (Istituto Nazionale di Astrofisica) at the
Spanish Observatorio del Roque de los Muchachos of the Instituto de
Astrofisica de Canarias. 

This research has made use of the Simbad data base, operated at CDS,
Strasbourg (France) and of data products from
the Two Micron All 
Sky Survey, which is a joint project of the University of
Massachusetts and the Infrared Processing and Analysis
Center/California Institute of Technology, funded by the National
Aeronautics and Space Administration and the National Science
Foundation. 

\end{acknowledgements}

\end{document}